\documentclass[prl,floatfix,twocolumn,showpacs]{revtex4}
\usepackage{amsmath}
\usepackage{amsfonts}
\usepackage{mathrsfs}
\usepackage{epsfig}
\usepackage{graphicx}
\usepackage{dcolumn}
\usepackage{expl3}
\begin{document}
%\title{Frequency switching induces control through spontaneous creation of (dynamic) chaotic traps.}
%\title{Creating chaotic traps through frequency switching can control dynamical systems}
%\title{Designer dynamics through chaotic traps: Controlling
%arbitrarily complex behavior in driven systems}
\title{Designer dynamics through chaotic traps: Controlling
complex behavior in driven nonlinear systems}
%\title{Chaos-induced order: Controlling dynamical systems by creating chaotic traps}
%through frequency switching}
\author{Shakti N. Menon$^1$, S. Sridhar$^{1,2}$ and Sitabhra Sinha$^1$}
\affiliation{
${}^{1}$\mbox{The Institute of Mathematical Sciences, CIT Campus, Taramani, Chennai 600113, India.}\\
${}^{2}$\mbox{Department of Chemistry, Brandeis University, Waltham, Massachusetts 02454-9110, USA.}\\
}
\date{\today}
\begin{abstract}
Control schemes for dynamical systems typically involve stabilizing
unstable periodic orbits. In this paper we introduce a new paradigm of
control that involves `trapping' the dynamics arbitrarily close to any
desired trajectory. This is achieved by a state-dependent dynamical
selection of the input signal applied to the driven nonlinear system.
An emergent property of the trapping process is that the signal
changes in a chaotic sequence: a manifestation of
chaos-induced order. The simplicity of the control scheme makes it
easily implementable in experimental systems.
\end{abstract}
%87.19.Hh Cardiac dynamics
%87.19.lr Neuroscience: Control theory and feedback
%05.45.-a Nonlinear dynamics and chaos
%05.45.Gg Control of chaos, applications of chaos

%=============================================================================%
%=============================================================================%

\maketitle

%%%% 1st paragraph %%%%

%The application of external stimuli to nonlinear dynamical systems can
%result in a rich repertoire of behavior
%External signals can elicit a rich repertoire of behavior in
%nonlinear dynamical systems that are often vital in the context of
%physiology
%The application of external stimuli to nonlinear dynamical
%systems can elicit a rich repertoire of behavior, often vital in
%physiological contexts
%Nonlinear systems perturbed by external signals can display a rich
%repertoire of behavior (exhibit novel dynamical regimes) that may differ
%significantly from their
%autonomous dynamics (behavior).
%Nonlinear dynamical systems 
%can respond to external signals by displaying
%a rich repertoire of behavior,
%When nonlinear systems are perturbed by external signals,
%they exhibit a rich variety of dynamical regimes that may differ
%significantly from their autonomous behavior~\cite{Kurths01}.
Nonlinear systems perturbed by external signals exhibit a rich variety
of dynamical regimes~\cite{Kurths01,Glass2001}.
%Such transitions can have vital consequences for the functioning
%of many natural systems~\cite{Glass2001}. Striking 
Examples 
%in the physiological context 
include
the entrainment of mammalian circadian rhythms to the day-night
cycle~\cite{Reppert2002} and the periodic cardiac contractions driven by
electrical signals from the
sinus node~\cite{Irisawa93,Rosen2011}. 
%Perhaps more spectacularly,
External signals 
%can sometimes 
may also
induce pathological conditions, 
such as 
when a precisely
timed electrical pulse triggers life-threatening arrhythmia in the
heart~\cite{Winfree80} or when aberrant sensory inputs
result in reflex epilepsy~\cite{Puranam2001}.
A deeper understanding of the dynamical principles governing the
response of the system to stimuli can therefore aid in the development
of more effective treatment of such disorders.
%The treatment of such disorders can potentially benefit from a deeper
%understanding of the dynamical principles governing the response of
%the system to stimuli {(\bf change?)}.
%of interaction between nonlinear dynamical systems and external
%signals
Indeed, 
%this assumes importance in view of
%the increasing recognition that the application of 
externally applied stimuli, such as electromagnetic pulses, 
have been used to treat
%can aid in the treatment of 
several clinical
disorders~\cite{George03,Luther2011}.
%although the operating principles are yet to be fully understood.
For example,
%Most notably, 
deep brain stimulation (DBS)~\cite{Kringelbach07}
has been applied 
%A particularly successful
%emerging therapy of this nature, 
%example is deep brain stimulation (DBS)~\cite{Kringelbach07}
%which, 
%although its operational principles are yet to be fully
%understood, has been
%used to treat 
to treat a range of ``dynamical diseases''~\cite{Glass77}, such as
Parkinson's disease~\cite{Rosin11}, major depressive
disorder~\cite{Warner-Schmidt13} and dystonia~\cite{Vidailhet2005}.
Therapies that involve external
electrical~\cite{Berenyi2012} or magnetic~\cite{George03} stimulation
are non-invasive, making them an attractive treatment option.
%for the treatment
%of depression, epilepsy and other disorders.
%The non-invasive nature of therapies involving external
%electrical~\cite{Berenyi2012} or magnetic~\cite{George03} stimulation
%make them an attractive option for the treatment of disorders such as
%depression and epilepsy. 
%Other
%non-invasive therapies such as trans-cranial magnetic~\cite{George03} and
%electrical~\cite{Berenyi2012} stimulation have been used to treat 
%disorders
%such as depression and epilepsy, respectively.

The theory of the control of dynamical systems via tunable
signals~\cite{Scholl08} provides a potential framework for
understanding the mechanisms underlying such
therapies.
%A potential framework for understanding the mechanisms underlying such
%therapies is provided by the
%theory of the control of dynamical systems via tunable
%signals~\cite{Scholl08}.
The dominant paradigm for the control of nonlinear systems
involves applying feedback
to perturb an accessible system variable or parameter to stabilize
an unstable periodic orbit (UPO)~\cite{OGY90,Boccaletti2000}. 
This approach
%pioneered by
%The dominant paradigm of control is the feedback scheme developed by
%Ott, Grebogi \& Yorke (OGY)~\cite{OGY90}, 
%which involves the application of small perturbations to an accessible system parameter. Several variations of this scheme has since 
has been implemented in 
diverse experimental
systems~\cite{Ditto90,Singer91,Roy92,Pierre96,Garfinkel92,Schiff94}.
%experiments on systems as diverse as magnetoelastic
%ribbons~\cite{Ditto90}, hydrodynamic flows~\cite{Singer91},
%lasers~\cite{Roy92}, ionization waves in plasma~\cite{Pierre96}, cardiac tissue~\cite{Garfinkel92} and {\it in vitro}
%neuronal populations~\cite{Schiff94}. 
%The efficacy of these schemes is not
%restricted to the control of chaotic dynamics alone. For instance, 
%The mechanism of control for all OGY inspired schemes involve using feedback
%to perturb a system variable or parameter, and thereby stabilize an unstable
%periodic orbit (UPO). 
Practical implementations of such ``closed-loop'' feedback schemes
require significant
computational effort as
%to continuously monitor the system and update the
%control signal. More generally, 
they typically presume
detailed knowledge of the dynamical state of the non-linear system
and often involve
high information processing overhead.
%as the control may be sensitive to small variations in the parameters.
%several parameters require  tuning. 
A fundamental limitation of such approaches is that the
permissible controlled states
%related as they are to existing UPOs, 
are intrinsically connected to the specific dynamics of the system.
From this perspective, a control scheme designed to
maintain the system around an arbitrary state, using minimal information
about its dynamics, will
%drive the system
%to accurately follow a
%``designer" 
%pre-designated 
%given trajectory of arbitrary complexity will
%be an important contribution in this area.
not only be a theoretical breakthrough, but would also widen the
horizon of control in practical applications.
%Indeed, the scope of such a method extends beyond the development of
%therapies for dynamical diseases discussed above. 	
%In principle, 
Such a method could be used to target
instabilities in any 
physical system driven by external signals, e.g., those observed in
magnetically confined plasma~\cite{Boozer2003}.
%, can be potential targets for the method we outline here. 

%%%%%%%3rd para%%%%%
In this paper, we present a novel paradigm for the
control of driven dynamical systems by restricting trajectories to a small but finite volume of phase
space.
%to user-defined ``designer'' states.
%through the application of external stimuli. 
%by switching between a specified set of values of an external signal.
%driving the system.
%The principle of the proposed approach 
%Its target is the 
%by switching between a specified set of values of the external signal.
This outcome is achieved with minimal information about the
state of the system, by switching between a specified set of values of
an external signal every time the state crosses a reference threshold.
Note that this reference state is quite distinct from the notion of a ``target'' state (e.g., as
in proportional feedback schemes~\cite{Astrom2010}) around which the system
is desired to be confined, and consequently the amplitude of the control signal need not be
modulated as a function of the difference between the instantaneous and target states. 
Consider the general setting of a dynamical system ${\cal
F}$ driven by an external signal $\Omega$:
%Central to our scheme is the simple switching algorithm
\begin{equation}\label{eq1}
X_{n} = {\cal F}(X_{n-1}, \Omega_{n})\,,\;\; \Omega_{n}={\cal
G}(X_{n-1})\,.
\end{equation}
Here $X_{n}$ is the state of the system at time
$n$, ${\cal G}$ is the deterministic control function, and the
signal $\Omega_{n}$ is 
%an externally accessible parameter of the system that is 
varied over time in order to impose control.
While ${\cal G}$ could be continuous, we demonstrate
that a suitably chosen discrete-valued function, that is
characterized by $\Omega$ switching between
a finite set of values, dynamically creates a trapping region around
a desired state of ${\cal F}$.
%For a suitable choice of ${\cal G}$, the application of 
%this scheme creates a trapping region around
%the desired state. 
On entering the switching-induced ``trap'', $X$ is
constrained to remain within it and follows a {\em chaotic}
trajectory,
even for regimes where the uncontrolled dynamical system ${\cal F}
(X_n, \Omega_n = \Omega)$ does not
display chaos. 
%Furthermore, although ${\cal G}$ is a purely deterministic rule,
%the sequence of $\Omega$s that results in control is also chaotic.
Furthermore, the sequence of $\Omega$s arising from the deterministic
rule ${\cal G}$ during control is also chaotic.
%control sequence $\{\Omega_1, \Omega_2, \ldots, \Omega_n\}$ is also
%chaotic even though ${\cal G}$ is a purely deterministic rule. 
Thus, chaos, which most control schemes try to eliminate,
is here an intrinsic feature of the controlled state. 
This outcome can hence be seen as a manifestation of ``chaos-induced
order'', analogous to the emergence of order in
stochastic systems through the application of
noise~\cite{Yamazaki98,Shinbrot2001}.
%The state itself maybe a trajectory
%Within this switching-induced ``trap'', the trajectory of $X$
%is {\em chaotic} even for regimes where the unperturbed dynamical
%system does not display chaos.
%Significantly, while ${\cal G}$ can in principle be continuous, 
We demonstrate
that a binary-valued function ${\cal G}$ involving discrete switching between 
$\Omega^a$ and $\Omega^b$ is sufficient to
control ${\cal F}$. 
The efficacy of our scheme is highlighted using two dynamical systems
%that describe phenomena mediated through 
driven by external signals. In each case, we show
that by starting with a sufficiently large trapping region and
strategically altering $\Omega_n$, the dynamics can be constrained to
any desired phase space volume.  
While the idea of trapping regions have been used earlier in very
specific contexts, such as the modulation of
mixing~\cite{Shinbrot93},
our method creates an adaptable trap that can control
%allows a trap to be dynamically altered in order to control 
the system around an arbitrary state.

%%%%%%%%%4th para%%%%%%%%%%
We first consider a generic model
for complex multi-periodic behavior, used to describe the dynamics in
systems as diverse as
the heart~\cite{Glass91} and magnetically confined
plasma~\cite{Boozer2003}, viz., the circle map (also known as the
Chirikov-Taylor map):
%simple dynamical system, the circle map (also known as the
%Chirikov-Taylor map), a generic model
%for complex multi-periodic behavior used to describe 
%systems as diverse as
%the heart~\cite{Glass91} and magnetically confined
%plasmas~\cite{Boozer2003}: %Winfree80,
\begin{equation}
\theta_{n+1} = \theta_{n} + \Omega_{n} -
k\,\sin(2\,\pi\,\theta_{n})\;\;\;\; (\mbox{mod}\;1)\,.
\label{eq2}
\end{equation}
Here, $\theta_{n}$ is the phase angle describing
the state of the system at time $n$, while $\Omega_{n}$ is
the corresponding driving frequency.
The value of $k$ determines the curvature of the map and thereby, the
periodicity of the dynamics. For all simulations reported here, we use values
of $k \leq 1$, a regime in which the map is invertible and the dynamics
are consequently non-chaotic.
We control the system by ``trapping'' the trajectories around a
desired value $\theta_{0}$, an outcome that we achieve by 
switching between two driving frequencies
$\Omega^a$ and $\Omega^b$ in a dynamically evolving sequence.
%in order to ``trap'' the trajectories around a desired value $\theta_{0}$. 
The choice of $\Omega$ at any instant $n$
depends on the sign of $\theta_{n}-\theta_{0}$.
Specifically, we set $\Omega_{n}=\Omega^{a}$ for $\theta_{n}<\theta_{0}$ and
$\Omega_{n}=\Omega^{b}$ for $\theta_{n}\geq\theta_{0}$.

%=============================================================================%
\begin{figure}
\begin{center}
\includegraphics[width=0.99\linewidth]{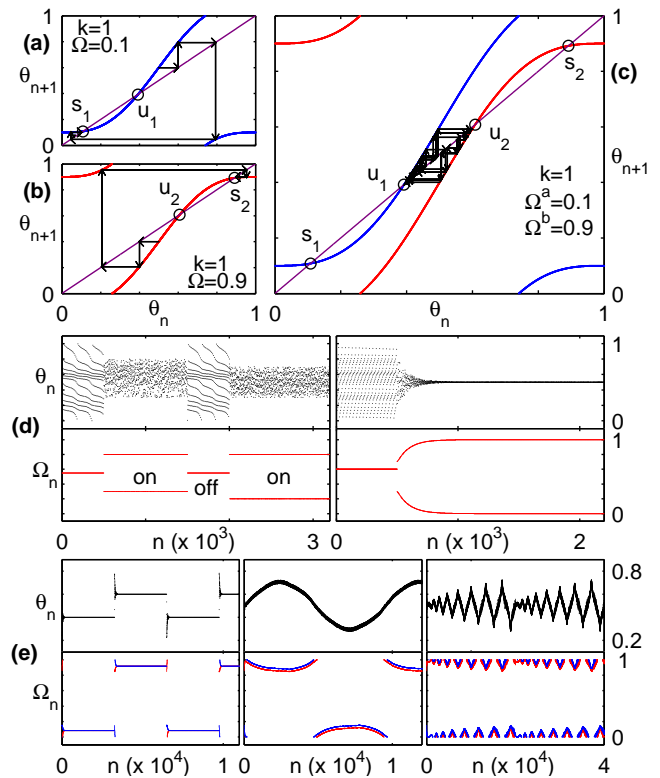}
\end{center}
\caption{
Demonstration of control in the context of the circle map
by switching between the two maps
%Controlling the dynamics of the circle map.
%(a-b) Cobweb plots of the trajectories in the specific maps
(a) $F_1$: $k=1$, $\Omega=0.1$ and (b) $F_2$: $k=1$, $\Omega=0.9$,
resulting in the controlled situation shown in (c).
%(c) Cobweb plot of the controlled situation obtained by switching between
%$F_1$ and $F_2$ following the mechanism described in the text. 
The switching between $\Omega^a = 0.1$ and $\Omega^b = 0.9$
``traps'' the dynamical state in a region bounded between the unstable fixed
points $u_{1}$ and $u_{2}$ of the maps $F_1$ and $F_2$ respectively, which is
located away from the stable fixed points $s_1$ and $s_2$ of the two maps. 
%Cobweb plots of the regular circle map, with initial conditions
%$\theta_{0}=0.5$,
%are shown for the cases (a) $k=1$, $\Omega=0.1$, and (b) $k=1$,
%$\Omega=0.9$, where the arrows represent the
%iterative trajectories. (c) When we impose the condition that the
%choice of $\Omega$ is dependent on the
%sign of $\theta_{n}~-~\theta_{0}$, the trajectory remains trapped
%between the unstable fixed points, $u_{1}$ and $u_{2}$, of the
%constituent maps.
(d) Even in the absence of fixed points, control can be achieved (in
the sense of reducing the range of values over which $\theta$ can
dynamically vary) immediately
upon turning ``on'' the switching, i.e., alternating
$\Omega_n$ between a pair of values that characterize two maps
(left). It is possible to reduce the trapping region to an arbitrarily small
domain by strategically changing $\Omega_n$ and hence the maps between which
switching occurs (right). For both panels, the switching sequence $\Omega_n$
is shown underneath the time-series of the dynamical variable $\theta$ 
($k=0.9$).
%(d) In the left panel, the consequence of a simple ``on-off'' control
% sequence of $\Omega_{n}$ is displayed, while the right panel shows the
%effect of gradually changing $\Omega_{n}$ such that the variance in
%$\theta_{n}$ is made arbitrarily small.
(e) Demonstration of the generality of the control method by
stabilizing a variety of trajectories through appropriate switching
strategies:
%This control method is extremely general, allowing for the stabilization of
%a wide variety of trajectories, through appropriate switching strategies:
(left) rectangular wave, (center) sine-like wave and (right) a bursting pattern.
For each panel, the switching sequence $\Omega_n$ is shown underneath
the time-series of the dynamical variable $\theta$ ($k=0.9$).
}
\label{fig1}
\end{figure}
%=============================================================================%

%%%%%%%%% Description of Figure 1 %%%%%%%%%%
We first demonstrate the trapping scheme for the
%The 
situation in which the driving frequency, $\Omega$, of the circle map
permits the existence of stable and unstable fixed points.
%, allows for a
%succinct demonstration of our trapping scheme, as we show below. 
Fig.~\ref{fig1}~(a) and (b) illustrate the characteristic dynamics
of the circle map for two such values of $\Omega$. In each case,
the trajectories tend towards the respective stable fixed points.
However,
switching between these two values of $\Omega$ in accordance with our control
mechanism results in the effective trapping of trajectories between
the unstable fixed points of the constituent maps, as shown in Fig.~\ref{fig1}~(c).
%It is important to note that the
%The successful 
The implementation of this scheme is not contingent
on the existence of fixed points: in the absence of a stable fixed point, the
circle map exhibits aperiodic dynamics that can be constrained by
switching between values $\Omega^a$ and $\Omega^b$, above and below the
original driving frequency [Fig.~\ref{fig1}~(d, left)].

Our scheme also allows for
the system dynamics to be confined to 
%an arbitrary small region of $\theta_n$. 
an extremely small volume of phase space around a desired state.
This can be achieved 
in the situation where the variation of $\Omega_n$ leads to the emergence of
unstable fixed points in the constituent maps [Fig.~\ref{fig1}~(c)].
In such cases, the system can be brought arbitrarily close to any
desired state
by decreasing
the absolute value of the difference between the successive values of $\Omega$, 
while strategically ensuring that the trajectory
never leaves the trapping region [Fig.~\ref{fig1}~(d, right)].
%It is to be noted that 
%Such an outcome occurs only when the variation of $\Omega_n$ leads to
%the emergence of unstable fixed points in the constituent maps
%[Fig.~\ref{fig1}~(c)].
The versatility of our control strategy is highlighted by the fact 
that it can be used to achieve trapping around
any arbitrary trajectory $\theta_n$ through appropriate manipulations of
$\Omega_n$, as illustrated in Fig.~\ref{fig1}~(e).

%=============================================================================%
\begin{figure}
\begin{center}
\includegraphics[width=0.99\linewidth]{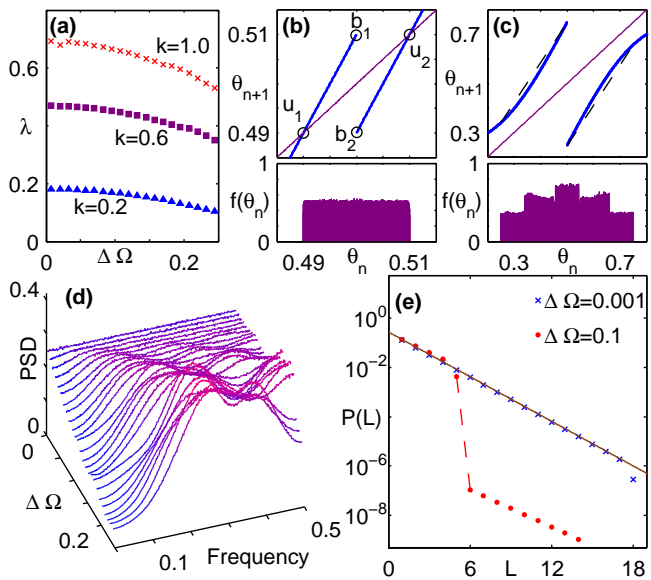}
\end{center}
\caption{
%Behavior for the situation where the two circle maps are characterized by
%$\Omega_{1}=1+\Delta\Omega$ and $\Omega_{2}=1-\Delta\Omega$.
Dynamical properties of the trajectories $\theta_{n}$ in the trapping
region, obtained by implementing the control scheme on the circle map,
using a switching sequence $\Omega_{n}=\{\Omega^a,\Omega^b\}$, are
illustrated for the case
$\Omega^a=1+\Delta\Omega$ and $\Omega^b=1-\Delta\Omega$.
(a)~Lyapunov exponents for the time series $\theta_{n}$,
obtained for a range of values of $\Delta \Omega$ for $k=0.2$ (filled upward
triangles), $k=0.6$ (filled squares) and $k=1$ (crosses). 
(b-c, top)~Composite maps for the cases where $k=1$ and (b)
$\Delta\Omega=0.01$;
(c)~$\Delta\Omega=0.25$. In (b), $b_1$ and $b_2$ refer to the values of
$\theta_{n+1}$ at the 
%tips 
extremities of the two segments
of the composite map, while $u_1$ and $u_2$ are
the unstable fixed points. Note that the trapping region in (b) closely
resembles the Bernoulli map, while the broken straight lines in (c)
are used to highlight the local
curvature of the map. (b-c, bottom)~Corresponding histogram for the
trajectories. (d)~Power spectral density for the switching sequence
$\Omega_{n}$, obtained over a range of values of $\Delta\Omega$ for $k = 1$.
(e)~Probability that the switching sequence {$\Omega_n$} between the
two constituent maps contains a sub-sequence of length $L$ in which
it is unchanged.
Results are displayed for $k = 1$, $\Delta \Omega = 0.001$
(crosses) and $\Delta \Omega = 0.1$ (filled circles). The broken line
indicates
the transition between the upper and lower branches. The solid line illustrates
the characteristic $1/2^L$ behavior expected from a fair coin toss.}
\label{fig2}
\end{figure}
%=============================================================================%

%%%%%%%%% Description of Figure 2 %%%%%%%%%%%
In order to characterize the complex dynamics in the trapped region, 
the Lyapunov exponent, $\lambda$, is measured for the time series $\theta_n$.
%%%%%Till here June 3rd 2014
For the situation displayed in Fig.~\ref{fig2}, we switch between two maps
corresponding to $\Omega^a = 1 - \Delta \Omega$ and $\Omega^b = 1 + \Delta \Omega$.
Fig.~\ref{fig2}~(a) shows that the dynamics in the trapped region is observed
to be chaotic for a wide range of $k$ and the control parameter
$\Delta \Omega$. To understand the
origin of chaos in this controlled state, we look at the composite map, i.e.,
the effective map obtained on application of the control. 
It comprises two segments separated by a discontinuity that determines
the location of the trapping region.
For small values of
$\Delta \Omega$, the trajectories span the domain more or less
uniformly, as shown in Fig.~\ref{fig2}~(b), a feature that becomes
increasingly pronounced as the constituent maps are brought closer together.
This can be understood from the fact that as $\Delta \Omega$ decreases, the
trapping region of the composite map increasingly resembles the
Bernoulli map, which produces a dynamical sequence indistinguishable from
a fair coin toss~\cite{Ford83}. Although the dynamics remain chaotic for
larger values of $\Delta \Omega$, the trajectories tend to preferentially
frequent certain regions, as shown in Fig.~\ref{fig2}~(c). This arises due to 
the
increased curvature of the composite map. The transition in behavior
is manifested in the corresponding power spectral densities (PSD) of the
sequence of driving frequencies, $\Omega_n$ [Fig.~\ref{fig2}~(d)]. For large
$\Delta \Omega$ there exist peaks in the PSD, indicating that it is more likely
to obtain a sub-sequence, comprising either $\Omega^{a}$ or $\Omega^{b}$, with a specific length.
As $\Delta \Omega$ decreases,
sub-sequences of all lengths $L$ can be observed with the probability expected
for the Bernoulli map ($\sim 1/2^{L}$). This is
confirmed by examining the distribution of sub-sequences of length $L$ for
different $\Delta \Omega$ [Fig.~\ref{fig2}~(e)]. 
For a low value of $\Delta \Omega$, the probability of observing large
sub-sequences is comparatively high.
%As $\Delta \Omega$ approaches zero, the probability
%of larger sub-sequences increases, as shown in Fig.~\ref{fig2}~(e). 
%that large sub-sequences can still be observed for higher values of
%$\Delta \Omega$, provided fixed points exist [see lower branch in
%Fig.~\ref{fig2}~(e)]. 
The likelihood of observing such sequences is
very small for higher values of
$\Delta \Omega$ [see lower branch in 
Fig.~\ref{fig2}~(e)] as they occur only when the trajectory
is close to one of the unstable fixed points.
%Note that for $\Delta \Omega = 0$, the PSD is a delta function (not shown in
%Fig.~\ref{fig2} (a)).

%=============================================================================%

%%%%%%% Discussion of control in the circle map %%%%%%%

The observation of chaos in the trapped region can be explained by the
fact that the switching control scheme effectively transforms an invertible
dynamical system into a non-invertible one. As our method constrains
the dynamics arbitrarily close to a desired trajectory in such traps,
it suggests a paradigm of control distinct from
schemes that involve the stabilization of UPOs. A necessary condition
for the creation of a chaotic trap is that the chosen pair of input
frequencies tend to drive the trajectories onto opposite
branches of the resulting composite map. In general, 
%the tips of the
the values of the two segments of the
composite map at the discontinuity 
provide the maximum bound of the
trapping region. In the situation where fixed points exist, the
creation of a trap is dependent on the condition that
$b_1$ lies below $u_2$ and $b_2$ lies above $u_1$ [Fig~\ref{fig2}~(b)].
In this case, the trajectories can neither enter nor leave the trap,
although the extent of the trap can itself be dynamically varied,
as illustrated earlier in Fig~\ref{fig1}~(d).
The fact that our scheme constrains the phase space of the dynamics
through chaotic switching suggests that it is a mechanism analogous to
`noise-induced order'~\cite{Sagues2007}. 
In contrast to most feedback methods, which
require precise information about the present state of the system,
our scheme only depends on whether $\theta_n > \theta_0$ or not. 
One practical consequence of this is that the proposed control scheme is more
robust to inaccuracies in the measurement of the dynamical state.

\begin{figure}
\begin{center}
\includegraphics[width=0.99\linewidth]{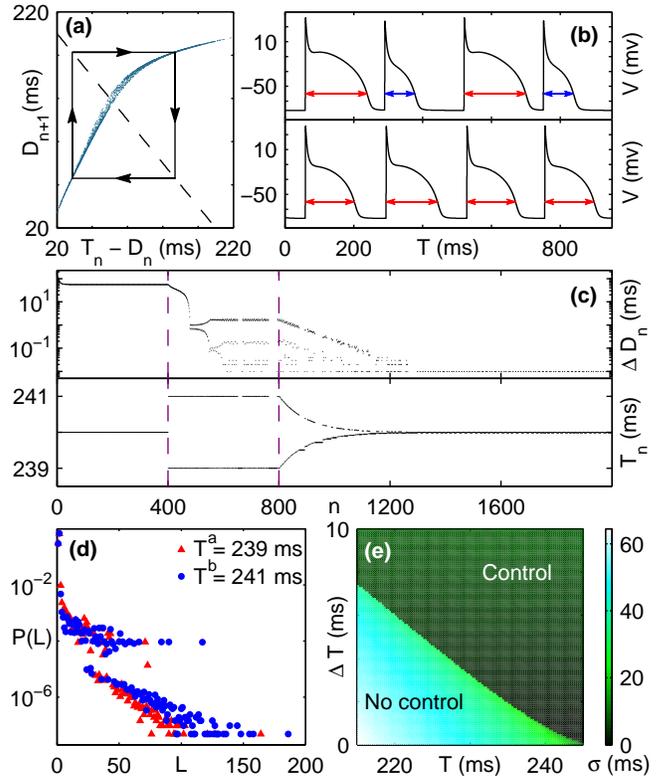}
\end{center}
\caption{Dynamics and control in the single-cell Luo-Rudy model.
(a) Restitution curve describing the relation between successive
action potential durations (APD) obtained using randomly chosen pacing 
periods, $T_n$. 
The trajectory of APD alternans, obtained for a 
pacing cycle length $T = 220$~ms (represented by a broken line), 
is indicated by arrows.
%For pacing cycle length $T = 220$~ms the system exhibits Period-2
%alternans in the APD, the corresponding trajectory being indicated 
%Period-2 alternans in the APD for the case $T = 220$~ms are indicated by
%arrows, with the pacing cycle length, $T$, represented by a broken line. %$T-(T_n-APD_{n}$.
(b, top) The time series of voltage $V$ for the case of pacing with a period
$T=230$~ms, displaying alternating long and short APDs.
(b, bottom) Application of switching control to the above system 
with $\Delta~T=4$~ms 
results in almost equal APDs.
%(b) Restitution curve for the Lou-Rudy model, obtained using a
%range of random pacing periods, $T$. The system exhibits alternans
%behavior, as illustrated by the direction of arrows, shown for the case
%$T=220$ ms (where the dashed line corresponds to $CL-DI_{n}$).
(c) Difference of successive APDs ($\Delta D_n = D_{n+1} - D_n$) (top) as a function of the pacing interval (bottom). The pacing interval corresponding to the
initial alternans state is $T=240$~ms. Control is applied with $\Delta
T=1$~ms between $n = 400$ and $n = 800$. For $n > 800$, the value of
$\Delta T$ is decreased gradually to $0.01$ ms (bottom), resulting in a
marked reduction in alternans (top).
%decrease in the variation of successive APDs (top). 
(d) Probability that the switching sequence
of periods for the case of control with $T=240\pm1$~ms contains sub-sequences
of either $T^a=239$~ms (filled triangles) or $T^b=241$~ms (filled circles) of
length $L$. %for which the same pacing interval is maintained for consecutive stimulations.
(e) Efficacy of the control scheme in reducing alternans, measured through
$\sigma$, the standard deviation of the APDs, for different values of
$T$ and $\Delta T$.
%Control is characterized by
%the transition from period-2 to aperiodic switching.  
% Power spectral density of the switching sequence for the case $T=240\pm \Delta T$ ms,
%obtained for a range of values of $\Delta T$.
}
\label{fig3}
\end{figure}

Finally, to illustrate the generality of our scheme,
we use
%have also applied 
it to control the dynamics of a realistic model
of an excitable biological system, namely the Luo-Rudy (LR)
model~\cite{LR91} of an externally stimulated
cardiac ventricular cell.
%a ventricular cell, subjected to external stimulation.
The system exhibits a characteristic {\em action potential} behavior
in the cellular transmembrane potential difference $V$, i.e., a
stimulation above a given threshold leads to a large transient
excursion from the resting state~\cite{Sinha2015}.
%It is an example of an excitable medium which, on being stimulated
%above a given threshold, exhibits a large deviation from the resting
%state before returning to it after a recovery period~\cite{Sinha2015}. This
%characteristic behavior, known as an action potential, is seen for the
%cellular transmembrane potential difference $V$ in this model.
%To highlight the versatility of our scheme, we apply it on a model of 
%a ventricular cell. 
%We use the Luo-Rudy (LR) model~\cite{LR91} which belongs to a class
%of biologically excitable systems that, for a single cell, has the
%form:
%\begin{equation} \label{eq3}
%        \frac{dV}{dt}    = -\frac{I_{\rm{ion}}(V,g_i)}{C_m} + I_{\rm{app}}(t)\,,
%\end{equation}
%where $V$ (mV) is the potential difference across the cellular
%membrane, $C_m$ ($\mu$F cm$^{-2}$) is the transmembrane capacitance,
%$I_{\rm{ion}}$ ($\mu$A cm$^{-2}$) is the total current density
%through ion channels on the cellular membrane, and $g_i$ describes
%the dynamics of gating variables for the different ion channels.
%The equations are solved using the forward Euler scheme, with time
%step $dt=0.01$ ms.
The action potential duration (APD)
for the $n^{\rm th}$ stimulus, $D_n$, is measured as the time 
during which $V$ is above the threshold. 
%that the voltage spends above the threshold value of $V=-60$mV. 
%Cardiac muscle cells are capable of exhibiting pathological
%behavior characterized by alternating long and short
%action potentials~\cite{Hall97}.
%Such {\em alternans} arise from 
%the rapid stimulation (pacing) of the medium with a period $T$~\cite{Hall97}.
%the nonlinear property of restitution
%in cardiac tissue.
The duration of successive excitations depends on whether the
medium has been able to sufficiently recover from prior activation.
This is reflected in the {\em restitution curve} of
Fig.~\ref{fig3}~(a). Here, under periodic
pacing with period $T$, we observe that the duration of the $n^{\rm th}$
action potential increases with the magnitude of $T - D_{n-1}$.
%difference between $T$ and the $n-1^{\rm th}$ action potential duration. 
This nonlinear restitution property of cardiac tissue can result in
{\em alternans}: pathological
behavior characterized by alternating long and short
action potentials~\cite{Hall97}.
As seen in Fig.~\ref{fig3}~(b, top), rapid stimulation of a single
cell can give rise to such a period-2 response in APDs.
%The nonlinearity of the restitution curve can
%give rise to a period-2 response in APDs on very rapid stimulation. 
%Fig.~\ref{fig3}~(b, top) shows the profile of the action potentials
%for such an
%instance of APD alternans, obtained by rapidly stimulating a single cell with
%a time period $T = 230$ ms. 
%Here, the steady state corresponds to the APDs
%alternating between long and short values. 

%=============================================================================%
%=============================================================================%

In order to control this behavior in a single cell, we
alter the pacing period by an amount $0 < \Delta T \ll T$, such that the period
is either $T^a = T - \Delta T$ or $T^b = T + \Delta T$, depending on the sign
of $D_n - D_{n-1}$. 
Varying the pacing periods in this manner
%which when applied individually still produce alternans,
maintains a state characterized by near-identical APDs.  
Note that (i) $\Delta T$ is chosen such that pacing exclusively with
either $T^a$ or $T^b$ produces alternans, and that (ii) switching between $T^a$ 
and $T^b$ is equivalent to alternating between a pair of values
of $\Omega$ in the system~(\ref{eq1}). 
%For more details of the LR model, refer to the supplementary
%material.  
In Fig.~\ref{fig3}~(b, bottom), we show the result of applying the
control using $\Delta T = 4$ ms (i.e., stimulating the cell with
pacing periods $T= 230\pm 4$ ms) on the alternans state
shown in Fig.~\ref{fig3}~(b, top). The system quickly reaches a state
where the APDs have almost 
equal values, indicating the effective termination of alternans [Fig.~\ref{fig3}~(c)].
The irregular nature of the switching dynamics is indicated by the wide distribution of
lengths $L$ of sub-sequences containing either $T^a$ or $T^b$ [Fig.~\ref{fig3}~(d)].

We quantify the efficacy of the switching control scheme 
through the standard deviation,
$\sigma$, of the resulting APD time-series
[Fig.~\ref{fig3}~(e)]. In situations where the system exhibits 
alternans, $\sigma$ is large. 
When $\Delta T$ is increased above a critical value,
the value of this measure decreases sharply and 
%there exists a critical $\Delta T$ above which control can be
control is
achieved. 
As $T$ decreases, the restitution effect becomes more
pronounced and the critical value of $\Delta T$ necessary for control 
is larger.
%Furthermore, this critical value decreases with an increase in $T$.
The achievement of control is characterized by the transition from a period-2 %biperiodic ?
switching mechanism (alternating at every step between periods
$T + \Delta T$ and $T - \Delta T$) to an aperiodic stimulation sequence [as in Fig.~\ref{fig3}~(d)].
Thus, the onset of control is indicated by the emergence of aperiodic or
chaotic switching behavior, suggesting a
practical method for inferring whether control has been achieved in
%novel method for verifying the achievement of
%potential 
%{\bf rough and ready} method 
%indicator for the achievement of
%control in
{\em in vivo} experiments.     

%=============================================================================%

%{\bf Last para:}
This paper introduces a novel control paradigm with
relatively low information
processing overhead, utilizing threshold-based
switching between a set of input signals.
In contrast to conventional feedback control schemes,
%that is the cornerstone of feedback-based methods.
our method implicitly uses chaos to trap the system dynamics around a
desired trajectory of arbitrary complexity.
%arbitrarily close to a desired state. 
%We connect it to previous work on creating trapping regions in the context
%of mixing in dynamical systems~\cite{Shinbrot93}.
%Our scheme has control
%efficiency comparable or better than existing methods and has
%much less information processing overhead.
Such a dynamical intervention scheme
%can be used for constraining a driven system to
%exhibit surprisingly complex state transitions
%and 
can have several
potential clinical applications, especially in the context of
``electroceutical'' therapies~\cite{Famm13} and treating
cardiac
arrhythmias~\cite{Hall97,Christini2001,Christini2006,Handbook2007}.
%Extensions and limitations
A variant of the control scheme has been used to control cardiac
alternans {\em in vitro},
wherein the frequency of the applied stimulus is determined by the variation in the
measured ventricular pressure~\cite{Sridhar13}. 
%While we have described the principle of the control scheme using circle map,
The scope of this switching method can be expanded to
include control in spatially
extended systems, e.g., coupled circle map lattices,
making it experimentally realizable in systems like magnetically confined 
plasma and lasers.
%
%As all biological systems are subject to noise, it is vital that our
%control scheme
%is robust to fluctuations as discussed in~\cite{Sridhar13}.
%========================================
%{\bf SI}
%$C_m$(= 1 $\mu$F cm$^{-2}$)
%The maximum Ca$^{2+}$ channel
%conductance is set to $G_{si}$ = 0.09 mS cm$^{-2}$ so as to get
%period-2 alternans on very rapid external stimulation. 
%The external pacing stimuli is a current $I_{\rm{app}}$ of magnitude
%$80 \mu$A cm$^{-2}$ applied for a duration of $0.5$ ms,
%
%Figure
%Histograms for the behavior of the case $k=0.8$, are shown at different
%iteration numbers $n$, in the situations when (a) For $\Delta \Omega=0.1$,
%only those trajectories that start between the two fixed points remain in the
%trapping region, while the others approach one of the stable fixed points.
%(b) Starting with $\Delta\Omega=0.3$ and gradually decreasing it, all
%trajectories approach the trapping region and will remain there.
%
%=============================================================================%

We thank V. Balakrishnan, C.~K.~Chan, N.~G. Garnier, G.~I.~Menon, A.
Pumir and V. Sasidevan for
helpful discussions.
%C K Chan for experimentally verifying our predictions for alternans
%control.
This research was supported in part by the IMSc Complex Systems Project.
We thank the HPC facility at IMSc for
providing access to ``Satpura'' which is partly funded by DST
(Grant No. SR/NM/NS-44/2009).
%=============================================================================%
%=============================================================================%

\end{document}